\documentclass[aps,prd,showkeys,showpacs,nofootinbib,preprintnumbers,twocolumn]{revtex4-2}
\usepackage[ansinew]{inputenc} % ASCII (Western CP)
\usepackage{graphicx,epsfig}
\usepackage{url}
\usepackage{color}
\usepackage[colorlinks, allcolors=black]{hyperref}
\usepackage{float}
\usepackage{bm}
\usepackage{dcolumn}
\usepackage{amsfonts,amssymb,amsmath}
\usepackage{psfrag}
\usepackage{subfigure}
\usepackage{natbib}
\usepackage{latexsym}
\begin{document}
\title{Scalar modes of polarization and speed of gravitational waves in $f(R)$  gravity }

%----------------------------------------------
%------------------------------------------------------------------------------------------------------------%

\author{Utkal Keshari Dash}
\email[]{cosmoskunal94@gmail.com}
\author{Bal Krishna Yadav}
\email[]{balkrishnalko@gmail.com}
\author{Murli Manohar Verma}
\email[]{sunilmmv@yahoo.com}

\affiliation{Department of Physics, University of Lucknow, Lucknow 226 007, India}
%--------------------------------------------------------------------------------------------------------------------------------------------------------%
%\date{\today}
\begin{abstract}
We explore the gravitational waves (GWs) within the framework of the $f(R)$ gravity model represented by $f(R)=R^{1+\delta}/R^\delta_c$ in the weak field approximation. In this scenario, gravitational waves exhibit an additional polarization mode beyond the standard transverse-traceless (TT) tensor modes. We show that the polarization characteristics of these waves are connected to the scalaron mass and the effective potential derived from the function $f(R)$.    Furthermore, the study of the speed of gravitational waves ($c_g$) within the Horndeski theory, particularly using the $f(R)$ model, reveals an intriguing feature about  the equality of  the speed of gravitational waves  to  that that of electromagnetic waves. This equivalence arises due to the modification introduced in the Ricci scalar within the $f(R)$  model.
\end{abstract}
%\keywords{Gravitational waves; scalaron; Longitudinal mode of Polarisation.}
%\pacs{04.80Cc, 04.50.Kd, 04.80.Nn}
\maketitle

\section{\label{1}Introduction}
Einstein's General Relativity (GR) has  evidently stood as  an invaluable tool for addressing a wide range of cosmological issues related to gravitation. We have amassed a wealth of observational evidence supporting the existence of dark energy, such as data from sources like Supernovae type Ia, Baryon Acoustic Oscillation (BAO), Cosmic Microwave Background (CMB)   anisotropies, and weak gravitational lensing, among others \cite{b1, b2, b3, b4}. Dark energy,  though not yet well known,  is  largely  understood to be   responsible for the  ongoing  accelerated expansion of the universe.

Efforts to modify GR can be broadly categorized into two classes: (i) modified gravity models and (ii) modified matter models. In  modified gravity models \cite{b11, b12, b13, b14, c1}, the gravitational aspect of the Einstein-Hilbert action is altered in various ways. Numerous modified gravity models, such as $f(R)$ theory, Scalar-Tensor theories, Braneworld models, and Gauss-Bonnet dark energy models, have been proposed and explored. Conversely, modified matter models \cite{b7, b8, b9, b10} introduce an additional component of matter into the energy-momentum tensor of GR. Quintessence \cite{b20}, k-essence, and phantom  dark energy models are some examples of modified matter models.

The standard model of cosmology, often referred to as the Lambda Cold Dark Matter ($\Lambda$CDM) model, provides a robust framework for explaining the late-time cosmic acceleration \cite{b5, b6, b16}. However, it faces challenges related to the cosmological constant $\Lambda$. Modified gravity models, particularly   $f(R)$  gravity models,  have emerged as successful alternatives for addressing both the dark matter and dark energy problems by leveraging the concept of the scalaron mass. This mass exhibits dark matter-like effects on local scales and behaves as dark energy on cosmological scales\cite{b01}.

Among the various $f(R)$  models, the present  paper focuses on our previous   model $f(R) = R^{1+\delta}/ R^{\delta}_{c}$ as discussed in  detail  in our  work at small and large scales \cite{b03, b05,  b06}. This model provides a sophisticated explanation of the two distinct phases of the universe. Specifically, it delves into how the mass of the scalaron depends on the $\delta$ parameter at galactic scales, revealing a sharp decrease for smaller values of $\delta$ and a subsequent increase for larger values. This insight enhances our understanding of the dynamical behaviour of the universe at  both,  local and cosmological,  scales.

The significant progress in addressing cosmological challenges owes much to the introduction of gravitational waves (GWs), a ground breaking discovery made in 2023  when the NANOGrav Collaboration \cite{b04} successfully detected gravitational waves through variations in Pulsar timing arrays \cite{i1}. Several  findings suggest that these waves hold tremendous potential for shedding light on various cosmological questions, including the early inflationary epoch,  where we have already shown the scalar spectral index and  tensor-to-scalar ratio being very close to the recent observational data \cite{b05},  age of the universe, the Hubble parameter, dark matter and dark energy. This also proves  to be an asset for detection of black holes (BHs), Super Massive Black Holes(SMBHs) and their  dynamical evolution such as spin etc \cite{g2}.

In the framework of Einstein's General Relativity (GR), the plane wave solution for gravitational waves encompasses two distinct polarization modes as  plus (+) mode polarization and cross($\times$) mode polarizations,  which shows the behaviour of gravitational waves in weak field limit in GR.

This paper aims to focus on  the gravitational waves in the weak field approximation limit in the modified $f(R)$ gravity model and investigate  the  modes of polarisation that exist due to the effective generalisation of the Ricci scalar. It has been argued that   the energy carried by these  additional modes would cause perturbation in the angular momentum of the inspiralling compact  binary system and change the phase evolution of tensor modes.  We hope that  if the phase correction due to scalar modes  can be measured through their  distinct waveforms, that would indicate the existence of these  modes and would be able to constrain $f(R)$ model parameters.  The observations  of GW170814  and GW170817 also put some constraints on  the energy flux of the scalar modes comparable in magnitude to that  of the tensor modes \cite{i3}.

	Another important aspect discussed  in this paper is the speed of gravitational waves $c_g$.  To understand the puzzling  features  of the universe like the events occuring beyond the event horizon of the black hole and supermassive black holes, pre-big-bang evolution, and  other unsolved  cosmological problems, we need to have a clear  knowledge of the speed of GWs along with its counterpart electromagnetic (EM) waves. From theoretical point of view, if the speed of GWs exceeds  the speed of EM radiations, then the GWs can show a distinct  behaviour as they  propagate  inside the highly dense objects like SMB or  as stochastic waves in post-inflationary era. Further,  if we find the speed as scale dependent,  it can also lead to significant features  at quantum scales too  \cite{g6}.

Using the more generalised Horndeski theory, we can estimate the speed of GWs  \cite{g5,g1}. There are various models like Quintessence and k-essence, Brans Dicke model, Covariant Galileons, Derivative couplings, Gauss-Bonnet couplings etc.  but in this paper we have used f(R) theory as a special case of Horndeski theory, to calculate the speed of gravitational waves and  compare it  to the  EM counterpart.

 Thus, the present  paper is organized in  four  sections. In Section II  we discuss the background of  field equations  in $f(R)$  gravity and conformal transformation from  the Jordan frame to the Einstein frame including an introduction of GWs in the field equations.  Section III  covers solution to  waves equations  in the $f(R)$  gravity model and examination  of the resulting  polarisation modes.  In  Section IV,   we explore  the speed of these waves in our specific  model. Section V encapsulates the summary and conclusion.

	\section{\label{2} $f(R)$ gravity and linearization of gravitational waves}
	 In this section,   we discuss about the modification made  in geometric part of the Einstein's field equations.
	
We begin with the 4-dimensional action in the general  $f(R)$  gravity model  which contains the modification of Ricci scalar in some functional form.  Thus,  the action is modified as
\begin{equation}
	\mathcal{A}=\frac{1}{2\kappa^2}\int d^x \sqrt{-g}f(R) + \mathcal {A}_{m}(g_{\mu\nu},\psi_m) \label{a1},
\end{equation}
where $\kappa^2 = 8\pi G$ and $\mathcal{A}_{m}$ is the action of the matter field $\psi_{m}$. The Friedmann-Lemaitre-Robertson-Walker (FLRW) spacetime is given as
	\begin{equation}
	\quad	ds^2=-dt^2+a^2(t)[dr^2+r^2(d\theta^2+sin^2\theta d\phi^2)] \label{a2},
	\end{equation}
where $a(t)$ is the time dependent scale factor and  speed of light $c$ = 1.
The Einstein's field equations then can be  obtained as
\begin{equation*}
	F(R)R_{\mu\nu}-\frac{1}{2}g_{\mu\nu}f(R)+g_{\mu\nu}\Box F(R)
\end{equation*}
\begin{equation}
\hspace{3cm}-\nabla_{\mu}\nabla_{\nu}F(R)=\kappa^2 T_{\mu\nu} \label{a3}, 
\end{equation}
where $F (R)= df/ dR $  and $T_{\mu\nu}$ is the energy-momentum tensor contributed by  matter.

Taking the trace of \textcolor[rgb]{0.00,0.00,1.00}{(\ref{a3})} in vacuum $(T_\mu^\mu = 0)$,    we obtain
\begin{equation}
	3\Box F(R)+RF(R)-2f(R)=0 \label{a4}   .
\end{equation}
By conformally transforming from the Jordan frame to the Einstein frame,   the  new  action can be written as \cite{bc1,b00,gw2}

\begin{equation*}
	\mathcal{A}=\int d^{4}x \sqrt{-\tilde{g}}\left[\frac{1}{2\kappa^2}\tilde{R}-\frac{1}{2}\tilde{g}^{\mu\nu}\partial_{\mu}\phi\partial_{\nu}\phi-V(\phi)\right] 
\end{equation*}
\begin{equation}
	\hspace{6cm}+\mathcal{A}_{m} \label{a5},
\end{equation}
where
\begin{equation}
	V(\phi)=\frac{FR-f}{2\kappa^2 F^2} \label{a6} .	
\end{equation}
represents the potential in a general class of $f(R)$ models showing the emergence of a  new scalar  degree of freedom
and an overhead tilde denotes the quantities in the Einstein frame.

 In this modified  gravity background,   we consider dark matter as scalaron particle corresponding to  field $\phi$ which is related to $F(R)$ in the Einstein frame as
\begin{equation}
	\Omega^{2}= F(R) = e^{-2Q\kappa\phi}\label{a7} ,
\end{equation}
where  $Q$  is   the strength of coupling between scalar field $\phi$ and non-relativistic matter  and   is given as
\begin{equation}
	Q=-\frac{1}{\sqrt{6}} \label{a8}.
\end{equation}

Thus, by  varying the action \textcolor[rgb]{0.00,0.00,1.00}{(\ref{a5})}  with respect to scalar field  $\phi$,  the  Klein-Gordon equation for effective scalar field $\phi$ is obtained as   \cite{b00,bc1,gw2}
\begin{equation}
	\tilde{\Box} \phi = \frac {dV(\phi)}{d\phi}+ \frac{\kappa}{\sqrt6}\tilde{T} \equiv  \frac{dV_{eff}(\phi)}{d\phi} \label{a9},
\end{equation}
where $V_{eff} (\phi)$  is the effective potential  whose second derivative provides scalar field a  mass as $m^2_\phi = d^2V_{eff}(\phi_{min})/{d\phi^2}$,  with $V_{eff} (\phi)$ having  the minimum value at $\phi_{min} $.

 The gravitational waves   are  extremely  weak  at the point very far from the sources,    and   so they can be linearized. Thus,  the metric can be characterized  in covariant form as  given below.
 \begin{equation}
 g_{\mu\nu}=\eta_{\mu\nu}+h_{\mu\nu} \label{a10}  .
 \end{equation}
 where $g_{\mu\nu}$ is the metric, $\eta_{\mu\nu}$ is the background metric and $h_{\mu\nu}$ is the small perturbation in the background metric with the condition $|{h_{\mu\nu}}|<<1$.

 The contravariant  counterpart  of the  metric in   \textcolor[rgb]{0.00,0.00,1.00}{(\ref{a10})}   can be written as
 \begin{equation}
 	g^{\mu\nu}=\eta^{\mu\nu}-h^{\mu\nu} \label{a11}.
 \end{equation}

The   linearization of   the modified Einstein  field  equations leads to

\begin{eqnarray*}
	F(R)R_{\mu\nu}+\frac{1}{2}(\eta_{\mu\nu}+h_{\mu\nu})f(R)+(\eta_{\mu\nu}+h_{\mu\nu})\Box F(R)
\end{eqnarray*}
\begin{equation}
		\hspace{4cm} -\nabla_{\mu}\nabla_{\nu}F(R)=0 \label{a12}.
\end{equation}
  It is assumed  that  the sources of  field for this condition  are   very far from the point where gravitational waves  are  detected, so $\mathcal{A}_m=0 \implies T_{\mu\nu}=0$ \cite{b00, i2}.

\section{Effect of scalaron mass in gravitational waves for  $ f(R) = R^{1+\delta}/  R^{\delta}_{c}$ type models}
 The constant tangential velocity condition of the test particle specifies that the motion of the test particle is stable in the orbits of spiral galaxies,  which indicates that   $f(R)$ is proportional to $R^{1+\delta}$,  where $\delta <<1$.
	A generic  $f(R)$   model gives rise to the scalar degree of freedom and, in particular,   can be  represented as
	\begin{equation}
	f(R)=\frac{R^{1+\delta}}{R^\delta_c} \label{a13},
	\end{equation}
	where $R_{c}$ is a constant having dimensions of Ricci Scalar $R$, $\delta$ is the model parameter and $\delta<<1$.
	The derivative of  $f(R)$  in   equation \textcolor[rgb]{0.00,0.00,1.00}{(\ref{a13})}   is given as
	\begin{equation}
		F(R)=(1+\delta)\frac{R^{\delta}}{R^\delta_c} \label{a14}.
	\end{equation}
	
	Now,  from  equation \textcolor[rgb]{0.00,0.00,1.00}{(\ref{a3})}, the field equations   can be written as
	\begin{equation*}
	G_{\mu\nu}\equiv\tilde{R}_{\mu\nu}-\frac{1}{2}g_{\mu\nu}\tilde{R}
	\end{equation*}
	\begin{equation*}
	=\frac{R^{\delta}_c}{R^\delta}\left[\frac{g_{\mu\nu}}{2}\left(\frac{R^{1+\delta}}{R^{\delta}_{c}}-(1+\delta)\frac{R^\delta.R}{R^\delta_c}\right)+\left({\frac{R^\delta}{R^\delta_c}}\right)_{,\mu,\nu}\right]
	\end{equation*}
 \begin{equation}
 	-\frac{R^{\delta}_c}{R^\delta}\left[g_{\mu\nu}\frac{\Box{R^\delta}}{R^\delta_c}\right]= 0  \label{a15}. 
 \end{equation}
 The  equation \textcolor[rgb]{0.00,0.00,1.00}{(\ref{a15})} further  leads to the following form
	\begin{equation*}
		G_{\mu\nu}=\tilde{R}_{\mu\nu}-\frac{\tilde{R}}{2}g_{\mu\nu}=\frac{R^{\delta}_c}{R^\delta}\left[\left({\frac{R^\delta}{R^\delta_c}}\right)_{,\mu,\nu}-g_{\mu\nu}\frac{\Box{R^\delta}}{R^\delta_c}\right]
	\end{equation*}
    \begin{equation}
    	-\frac{R^{\delta}_c}{R^\delta}\left[{\frac{1}{2}\eta_{\mu\nu}(\delta R^{\delta+1})}\right]=0  \label{a16}.
    \end{equation}
Recalling conformal transformation from   equation \textcolor[rgb]{0.00,0.00,1.00}{(\ref{a7})}
\begin{equation}
	\phi= \sqrt{\frac{2}{3}} \ln F \label{a17}.
\end{equation}
The Ricci scalar then gets transformed to
\begin{equation}
	R=R_{c}\left[\frac{e^{\sqrt{\frac{2}{3}}\kappa\phi}}{1+\delta}\right]^{\frac{1}{\delta}} \label{a18}.
\end{equation}

The modified Einstein's field  equations in the Einstein frame are  given by
\begin{equation*}
	\tilde{R}_{\mu\nu}-\frac{1}{2}g_{\mu\nu}\tilde{R} ={\sqrt{\frac{2}{3}}\kappa}\left[\partial_{\mu}\partial_{\nu}\phi-\eta_{\mu\nu}\Box{\phi} \right]
\end{equation*}
\begin{equation}
	-\frac{1}{2}\eta_{\mu\nu}\delta R_{c}\left[\frac{e^{\sqrt{\frac{2}{3}}\kappa\phi}}{1+\delta}\right]^{\frac{1}{\delta}}=0  \label{a19}.
\end{equation}

 As gravitational waves coupled with scalar field component propagate, the later gets perturbed  by a small change given  as $\Delta{\phi}$.
\begin{equation}
	\phi=\phi_{0}+\Delta{\phi}  \label{a20}.
\end{equation}
Using $h_{s}=\frac{\Delta{\phi}}{\phi_{0}}$,    we obtain
\begin{equation*}
	\tilde{R}_{\mu\nu}-\frac{1}{2}g_{\mu\nu}\tilde{R}={\sqrt{\frac{2}{3}}\kappa\phi_{0}}\left(\partial_{\mu} \partial_{\nu}h_{s}-\eta_{\mu\nu}\Box{h_{s}} \right)
\end{equation*}
\begin{equation}
		-\frac{1}{2}\eta_{\mu\nu}\delta R_{c}\left[\frac{e^{\sqrt{\frac{2}{3}}\kappa\phi}}{1+\delta}\right]^{\frac{1}{\delta}}=0  \label{a21}, 
\end{equation}
where $h_{s}$ is the massive mode polarization term which has been generated due to the scalar field $\phi$ which produces the longitudinal modes in traceless transverse modes.

Using equations  \textcolor[rgb]{0.00,0.00,1.00}{(\ref{a4})}- \textcolor[rgb]{0.00,0.00,1.00}{(\ref{a9})}  in previous Section,  we get
\begin{equation}
	m^2_\phi\left[\frac{3\delta(1+\delta)}{1-\delta}\right]=R_{c}\left[\frac{e^{\sqrt{\frac{2}{3}}\kappa\phi}}{1+\delta}\right]^{\frac{1}{\delta}}
\label{a22}, \end{equation}
  where $m_\phi$ term  represents   the mass of the scalaron and is given as \cite{b03}
  \begin{equation}
  		m^2_\phi=\frac{(1-\delta)^\frac{2\delta}{1+\delta}}{3\delta(1+\delta)}(R_c)^\frac{2\delta}{1+\delta}(\kappa^2\rho)^\frac{1-\delta}{1+\delta},
  \label{a23}\end{equation}
which also includes the source $T_{\mu\nu}$ whose trace is $\rho$.

Using equation \textcolor[rgb]{0.00,0.00,1.00}{(\ref{a22})}   in   equation \textcolor[rgb]{0.00,0.00,1.00}{(\ref{a21})}  we obtain
\begin{equation}
 \Box h_{s}= \sqrt{\frac{3}{2}}\frac{\delta^2(1+\delta)}{(\delta-1)\phi_{0}}m^2_\phi  \label{a24}.
\end{equation}

  From the above equation, it is evident that the massive mode polarisation depends on the mass of the scalaron which itself is dependent on $\delta$.  So,  the massive field shows complete dependence on the model.
  The plane wave equation for transverse mode of polarisation is written as
\begin{equation}
 	\Box h_{\mu\nu}=0 \label{a25}.
   \end{equation}
On solving equations  \textcolor[rgb]{0.00,0.00,1.00}{(\ref{a23})}  and   \textcolor[rgb]{0.00,0.00,1.00}{(\ref{a24})} together  we obtain  \cite{b00}
\begin{equation}
	h_{\mu\nu}= B_{\mu\nu}(\textbf{p})e^{ik^\alpha x_\alpha}\label{a26}
\end{equation}
and
\begin{equation}
	h_{s}= H(\textbf{q})e^{iq^\alpha x_\alpha}\label{a27}
\end{equation}  are  added  to  the plus and cross polarization terms are added,  which is now given as \cite{b00}
\begin{equation}
	\bar{h}_{\mu\nu} = h_{+} e^{(+)}_{\mu\nu} + h_{\times}e^{(\times)}_{\mu\nu}+ h_s e^{(s)}_{\mu\nu}  \label{a28}.
\end{equation}

Since the polarized wave is travelling  along  the Z-direction, these polarizations modify to

\begin{equation}
\bar{h}_{\mu\nu}=
\begin{bmatrix}
	
	1 & 0 & 0 & 0\\
	
	0 & h_+ & h_\times & 0\\
	0 & h_\times  & -h_+ & 0\\
	0 & 0 & 0 & h_s
\end{bmatrix}\label{a29}.
\end{equation}

 The group velocity of $h_s$ can be given as
 \begin{equation}
 	v_g  =  \frac{p}{\omega_m}\label{a30}
 \end{equation}
\begin{equation}
	v_g  =  \sqrt {1-\left(\frac{m_\phi }{\omega_m}\right)^2}\label{a31}, 
\end{equation}
where $\omega_m = \sqrt{m_\phi^2 + {p} ^2}$ is the angular velocity of the massive gravitational wave propagating in the  longitudinal direction. Thus, by using the constraints on the scalaron mass from cosmological evolution and consistent with dark matter in  our earlier work \cite{b03},  we can constrain $v_g$. It may be compared to the constraints  obtained on the velocity   through  the mass of gravitons  $m_g$  as  $\left(m_g/\omega_{orbital}\right)^2 < 0.003$ by consideration  of energy loss in form of emission of  massive gravitons   from  compact binary systems  \cite {h1,h2}. In our model,  using  equation   \textcolor[rgb]{0.00,0.00,1.00}{(\ref{a23})} with   \textcolor[rgb]{0.00,0.00,1.00}{(\ref{a31})},     the propagation velocity is given as 
\begin{equation}
v_g  = \sqrt{1-\frac{(1-\delta)^\frac{2\delta}{1+\delta}}{12 \pi^2 n^2\delta(1+\delta)}R_c^\frac{2\delta}{1+\delta}(\kappa^2\rho)^\frac{1-\delta}{1+\delta}}\label{a311}, 
\end{equation}
where $n$ is the wave frequency.   Clearly,   $v_g$  is sensitive to $R_c$,    $\delta$  for a given matter background   (both of which together determine the scalaron mass),  and  wave frequency depending on the dynamically evolving sources of GWs  (which includes a stochastic GW background).  This can be used as an effective   tool to distinguish the cosmological  models beyond GR  from the GR-based models  over a wide range of frequencies in the future   GW  detectors.  We explore  the behaviour of the velocity $v_g$ on our  model parameters as the waves propagate through the  matter background. 

%---------------------------------------------------------------
\begin{figure}[!h]
	\centering \begin{center}  \end{center}
	{{\includegraphics[width=0.44 \textwidth,origin=c,angle=0]{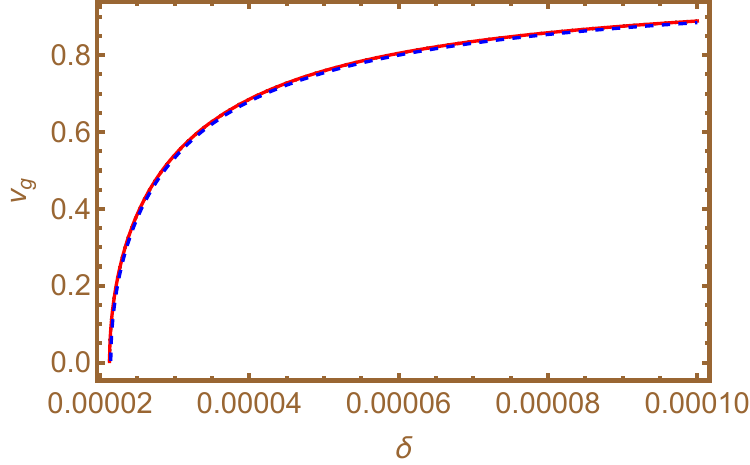} }}%
	%\qquad
	%%{{\includegraphics[width=0.44 \textwidth,origin=c,angle=0]{V_ss_1_P4.pdf} }}%
	\caption{Variation of  $v_g$ vs $\delta$ for  $R_c$ = $\Lambda$ (cosmological constant = $10^{-84} (GeV)^2$),  $10^{-78}(GeV)^2$  and $1(GeV)^2$,  shown as black dotted, red and blue dashed curves, respectively.  Value of energy density  of background  matter $\rho = 10^{-42}(GeV)^4 $  is chosen at the galactic scales  and frequency is chosen as the lowest  ($10^{-18}$ Hz) in the ELF band to see its comparable effect with respect to $R_c$  in the $v_g$.}%
	\label{f1}%
\end{figure}
%------------------------------------------------------------------------------------------------------
\begin{figure}[!h]
	\centering \begin{center}  \end{center}
	{{\includegraphics[width=0.44 \textwidth,origin=c,angle=0]{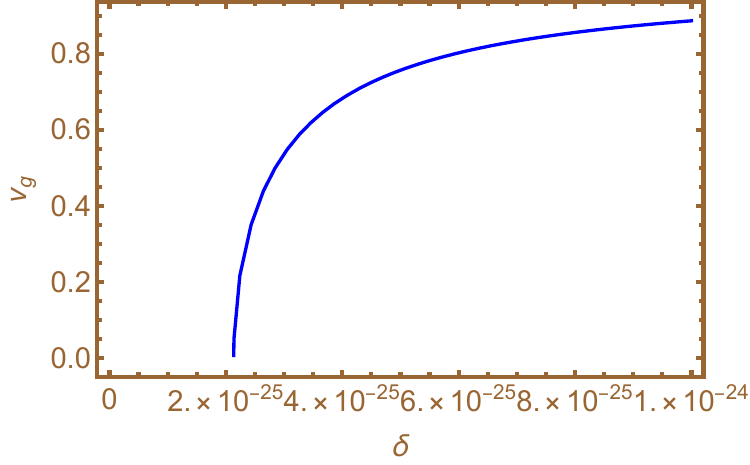} }}%
	%\qquad
	%%{{\includegraphics[width=0.44 \textwidth,origin=c,angle=0]{V_ss_1_P4.pdf} }}%
	\caption{Plot of    $v_g$ vs $\delta$    for  frequency  $=10^{-8}$Hz relevant to  NANOGrav band. Velocity  approaches $1$  for  $\delta  \sim 10^{-24}$.    $R_c$ = $\Lambda$ (cosmological constant = $10^{-84} (GeV)^2$) and   $\rho = 10^{-42}(GeV)^4 $.   }%
	\label{f11}%
\end{figure}
%----------------------------------------------------------------------------------------------------------------

\begin{figure}[!h]
	\centering \begin{center}  \end{center}
	{{\includegraphics[width=0.44 \textwidth,origin=c,angle=0]{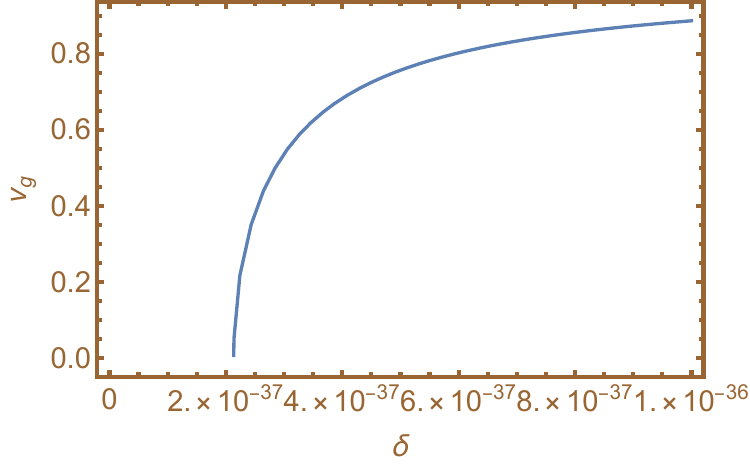} }}%
	%\qquad
	%%{{\includegraphics[width=0.44 \textwidth,origin=c,angle=0]{V_ss_1_P4.pdf} }}%
	\caption{Plot of    $v_g$ vs $\delta$  for  frequency = $10^{-2}$Hz  of the expected  range  of LISA.  Velocity  approaches $1$  for  $\delta  \sim 10^{-36}$.    $R_c$ = $\Lambda$ (cosmological constant = $10^{-84} (GeV)^2$) and   $\rho = 10^{-42}(GeV)^4 $. }%
	\label{f12}%
\end{figure}

%-----------------------------------------------------------------------------------------------------------------------

\begin{figure}[!h]
	\centering \begin{center}  \end{center}
	{{\includegraphics[width=0.44 \textwidth,origin=c,angle=0]{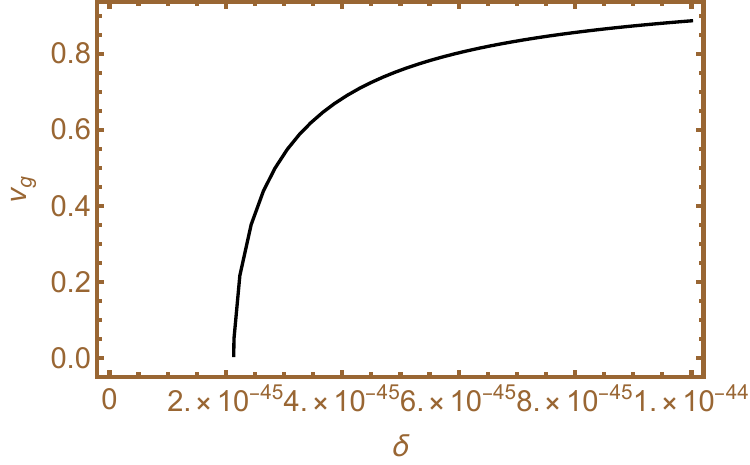} }}%
	%\qquad
	%%{{\includegraphics[width=0.44 \textwidth,origin=c,angle=0]{V_ss_1_P4.pdf} }}%
	\caption{Plot of    $v_g$  vs $\delta$ with   for  frequency =  $100$ Hz  of  LIGO window. Velocity  approaches $1$  for extremely low value of  $\delta  \sim 10^{-44}$.  $R_c$ = $\Lambda$ (cosmological constant = $10^{-84} (GeV)^2$) and   $\rho = 10^{-42}(GeV)^4 $.   }%
	\label{f13}%
\end{figure}
%-----------------------------------------------------------------------------------------
\begin{figure}[!h]
	\centering \begin{center}  \end{center}
	{{\includegraphics[width=0.44 \textwidth,origin=c,angle=0]{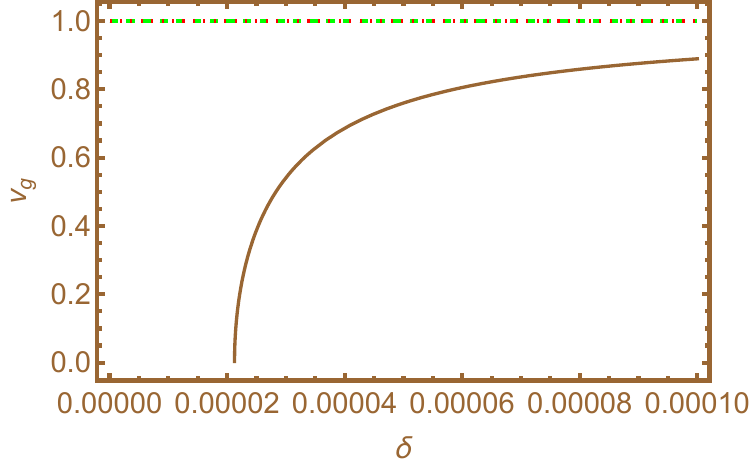} }}%
	%\qquad
	%%{{\includegraphics[width=0.44 \textwidth,origin=c,angle=0]{V_ss_1_P4.pdf} }}%
	\caption{Multi-frequency variation of  $v_g$ vs $\delta$ for  a set of  different frequencies.    Brown curve represents frequency $10^{-18}$ Hz, while the green dashed,  yellow dashed and red dashed curves correspond to the frequencies $10^{-8}$ Hz,  $10^{-2}$ Hz  and  $100 $ Hz,   respectively. $R_c$ = $\Lambda$ (cosmological constant = $10^{-84} (GeV)^2$)  and  $\rho = 10^{-42}(GeV)^4 $.  }%
	\label{f14}%
\end{figure}

%-------------------------------------------------------------------
\begin{figure}[!h]
	\centering \begin{center}  \end{center}
	{{\includegraphics[width=0.44 \textwidth,origin=c,angle=0]{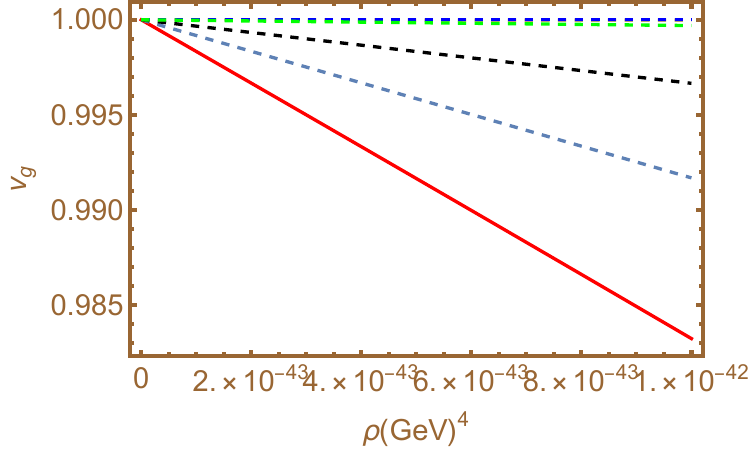} }}%
	%\qquad
	%%{{\includegraphics[width=0.44 \textwidth,origin=c,angle=0]{V_ss_1_P4.pdf} }}%
	\caption{This plot shows $v_g$ vs $\rho$ for $R_c$ = $\Lambda$ (cosmological constant = $10^{-84} (GeV)^2$).  Smooth red, deep rainbow, black, green and blue dashed lines correspond to  $\delta$= $2\times10^{-5}$,   $10^{-4}$, $10^{-3}$, $10^{-2}$ and  $0.1$,  respectively.  Frequency is chosen as the lowest  ($ 10^{-18}$Hz) in the ELF band. }%
	\label{f2}%
\end{figure}

%----------------------------------------------------------------

Figure   \textcolor[rgb]{0.00,0.00,1.00}{\ref{f1}}  describes the role of parameter $\delta$
in  variation  $v_g$  of    GWs. It shows a range of   $\delta$  up to an  order of $10^{-4}$.   This  speed  asymptotically  approaches  the  value of speed of light beyond this value of $\delta$.   This indicates that if the mass of the scalaron is heavier then it would more  strongly  impede   the  propagation of GWs through  galactic background     ($\rho = 10^{-42}(GeV)^4 $),  and  the longitudinal mode would  tend to disappear. This is a manifestation of the  chameleon mechanism in  the higher density regions where mass of scalaron  $m_\phi$ becomes too  large in response to the matter  environment   and consequently,  the Compton wavelength of  scalaron becomes extremely small. This leads to the collapse of scalar field and  the gravitational wave loses the  scalar longitudinal mode.  The plot also shows  that the speed  of the gravitational waves  does not change much  for different values of $R_c$.  This is because the value of $R_c$ is  extremely  small compared to the  much higher  contribution from   frequency (which is actually the lowest, as taken here)  in the  extremely low frequency (ELF) band   $10^{-15}- 10^{-18}$ Hz,   where gravitational wavelength is  $\pi$ times  the Hubble distance  producing  quadrupolar  anisotropies in CMB  radiation.  Of course, when we  choose still  higher frequencies  corresponding  to  Laser Interferometer Gravitational Wave observatory (LIGO), future  Laser  Interferometer Space Antenna (LISA),  NANOGrav etc. and carry out a  similar analysis, we find the  increasing independence from $R_c$.

In addition to frequency $10^{-18}$ Hz  taken  in   Figure  \textcolor[rgb]{0.00,0.00,1.00}{\ref{f1}},  further,    the variation of $v_g$ is plotted for a set of other  frequencies ($10^{-8}$ Hz,  $10^{-2}$ Hz  and  $100 $ Hz)    in Figures  \textcolor[rgb]{0.00,0.00,1.00}{\ref{f11}}, \textcolor[rgb]{0.00,0.00,1.00}{\ref{f12}},  and   \textcolor[rgb]{0.00,0.00,1.00}{\ref{f13}},  respectively.  Here,   we  use a single value of $R_c = \Lambda$ and   a common  matter density background of galaxies $\rho = 10^{-42}(GeV)^4 $,     and attempt to examine the effects on these vastly different frequencies which arise from different sources and fall within the  detection range of the present or future  detectors.  It can be clearly  seen that for  higher frequencies,     $v_g$  gets   more sharply  closer to $c=1$  for successively lower values of model parameter $\delta$. This combined  multi-frequency behaviour over the large  range  of $\delta= 0-10^{-4}$   is shown in  Figure  \textcolor[rgb]{0.00,0.00,1.00}{\ref{f14}}, where only   $10^{-18}$ Hz coming from the stochastic  gravitational wave background displays a distinct variation, while at all other chosen frequencies, velocity  $v_g$   is uniformly equal to that  of light.

In  Figure \textcolor[rgb]{0.00,0.00,1.00}{\ref{f2}}, we   observe    a significant variation in  $v_g$  with the background matter density  for different values of $\delta$.  When $\delta$ is of  very low order $\sim10^{-5}$,    the value of $v_g$ drops sharply  with matter density,    but when it increases further, then close to $\delta=0.001$  and  $0.1$,   the curve approximates    speed of light $c=1$ for the given range of $\rho$.   This is due to rising  mass of scalaron    with decreasing value of $\delta$.  We find that    $\rho$  is also  dominant for the value of $\delta < 10^{-3}$ or lower, and becomes successively  less effective  for the higher values of $\delta$. Thus, smaller  values of $\delta$, and  higher background densities,  both  provide a higher mass to scalarons,  and so,  both  are  favourable  for reducing the speed of the longitudinal mode.   At  the given  galactic background density, the degree of elimination of the longitudinal mode will grow for smaller values of $\delta$ and may escape detection. Thus, using the cosmological  constraints  on mass of scalaron from the consideration of  age of the universe etc.,    we can conclude that the longitudinal mode of GWs  does not propagate at the speed of light   when $\delta$ is  very small and  mass of scalarons  becomes  high.

\section{Speed of gravitational waves in  Horndeski theory using $f(R)$ gravity}
The detection of  gravitational waves from the event GW170817 provided  the possibility of determination of their  velocity.  Several  efforts,  such as in Horndeski theory,   emerged with an extended   scope of study  \cite{g4,g5,g1,gw01}. The Lagrangian from the generalised action
\begin{equation}
	A=\int d^4x \sqrt{-g}L  \label{a32} 
\end{equation}	
	can be written using this theory as  \cite{g4}
\begin{equation*}
		L = G_2(\phi,X)+G_3(\phi,X)\Box\phi +G_{4}(\phi,X)R +G_{4,X}(\phi,X)[(\Box\phi)^2
\end{equation*}
\begin{equation*}
	 - (\nabla_\mu\nabla_\nu\phi)(\nabla^\mu\nabla^\nu\phi)] + G_5(\phi,X)G_{\mu\nu}\nabla^\mu\nabla^\nu\phi
\end{equation*}
\begin{equation*}
	-\frac{1}{6}G_{5,X}(\phi,X)[(\Box\phi)^3-3(\Box \phi)(\nabla_\mu \nabla_\nu \phi)(\nabla^\mu \nabla^\nu \phi)
\end{equation*}
\begin{equation}
	+2(\nabla^\mu \nabla_\alpha \phi)(\nabla^\alpha \nabla_\beta \phi)(\nabla^\beta \nabla_\mu \phi)] \label{a33},
\end{equation}
where $G_{2,3,4,5}$ depend on $\phi$ and $X$,  and  $G_{i,\phi}\equiv {\partial G_i}/{\partial \phi}$ and $G_{i,X}\equiv {\partial G_i}/{\partial X}$  with
 \begin{equation}
	X = -\frac{1}{2}\nabla_{\mu}\phi\nabla^{\mu}\phi \label{a37}.
\end{equation}

Using $f(R)$  gravity model we can write  action    \textcolor[rgb]{0.00,0.00,1.00}{(\ref{a32})}  in terms of equation   \textcolor[rgb]{0.00,0.00,1.00}{(\ref{a1})}.  Thus,  the Lagrangian can be transformed accordingly with   the choice of $G_2, G_3, G_4, G_5$,      and  all these parameters take  the form

\begin{equation}
	G_2 = RF-f, \hspace{0.2cm}
	G_3 = 0,  \hspace{0.2cm}
	G_4 = F,  \hspace{0.2cm}
	G_5 = 0  \label{a34}.
\end{equation}
Using the equivalence principle,    we can transform the values of the above parameters in terms of $\phi$ which  leads to   the following equations, 
\begin{equation*}
	G_2 = \delta R^{1+\delta}\left(e ^ {\sqrt{\frac{2}{3}}\kappa \phi}\right)^{\frac{1+\delta}{\delta}},
	G_3 = 0,
	G_4 = e ^ {\sqrt{\frac{2}{3}}\kappa \phi},
\end{equation*}
\begin{equation}
  G_5 = 0   \label{a35}.
\end{equation}
The speed of the gravitational waves    $c_g$    propagating over  a cosmological matter   background depends on the scalar field which can be written in the generalised form using Horndeski theory as
\begin{equation}
	c^2_g = \frac{2G_4 - (\dot{\phi})^2 G_{5,\phi} - \dot{\phi}^2\ddot{\phi} G_{5,X}}{2G_4 - (2\dot{\phi})^2 G_{4,X} + (\dot{\phi})^2 G_{5,\phi} - H(\dot{\phi})^3 G_{5,X}} \label{a36},
\end{equation}
where  
\begin{equation}
X= -\frac{1}{2}[\dot{\phi}^2 - {\nabla\phi}^2]\label {371}
\end{equation}
 $H$ is the Hubble parameter.
Using equation \textcolor[rgb]{0.00,0.00,1.00}{(\ref{a35})}  in equation \textcolor[rgb]{0.00,0.00,1.00}{(\ref{a36})} we obtain 
\begin{equation}
		c^2_g = \frac{2G_4}{2G_4 - (2\dot{\phi})^2 G_{4,X}} \label{a38}.
\end{equation}
Further,  we consider the second term of the denominator of equation \textcolor[rgb]{0.00,0.00,1.00}{(\ref{a36})} as 
\begin{equation}
	G_{4,X} = \frac{\partial G}{\partial X} = \frac{\partial e^{\sqrt{\frac{2}{3}}\kappa\phi}}{\partial[\dot{\phi}^2 - {\nabla\phi}^2]}=0 \label{a39}.
\end{equation}
From equations  \textcolor[rgb]{0.00,0.00,1.00}{(\ref{a38})}  and \textcolor[rgb]{0.00,0.00,1.00}{(\ref{a39})},   we obtain $c^2_g =1$
This implies that the GWs would  propagate with the speed of light in this case.

\section{Discussion and conclusion}
In the    present    paper,   we investigated  the gravitational waves in an  $f(R)$  gravity model and found   that the conventional transverse     polarisation modes  (plus and cross) are accompanied by a massive polarisation mode along the Z-direction of the wave propagation,   which  is produced  due to the scalar degree of freedom associated with the modification in the geometrical part of the Einstein-Hilbert action.  The scalar  field seems to play   no direct  role    in the transverse modes  of polarisation,  but causes  the creation and propagation of this  massive longitudinal mode. This   $h_s$   mode  varies  with $m_\phi$,    and  thus,   is sensitive to even an extremely   small variation in the powers  of Ricci scalar.  The  velocity curves  plotted  in   Figures   show   that this longitudinal mode  of gravitational wave cannot propagate  when $\delta=0$.    However, depending on frequency ($10^{-18}$ Hz,      $10^{-8}$ Hz,   $10^{-2}$ Hz and  $100$ Hz  corresponding to  stochastic GW background imprinted on CMB,   NANOGrav, future LISA and present  LIGO detectors, respectively),  the speed of propagation in galactic matter background  picks up     sharply   and   jumps to speed of light as $\delta$ is switched on.  Thus, for example,   we find that  for a  $100$  Hz   astrophysical GW   longitudinal mode,  its   velocity jumps  very   sharply  to speed of light even   as $\delta$   slightly  increases  beyond  $10^{-44}$. 

Thus,  it can be interpreted that in GR limit  where mass is infinite, the longitudinal modes of vibration cannot be formed.
On the other hand,  when we calculate  the speed of the gravitational waves   using Horndeski theory, it is  found   exactly  equal to the speed of light. Using the $f(R)$   model as a special case in Horndeski theory,   we have   only the Ricci curvature in the   field equations which contributes to the conformal part, but when we consider the Riemannian  part we find   an inequality  in the speed of GWs and speed of  light. This anomalous speed of GWs  is due to the involvement of Weyl curvature tensor when $G_5$ becomes non zero. These consideration   allow us to formulate a Weyl criterion for anamolous speed of spin-2 GWs \cite{g6,g3} using the Newman-Penrose formalism which will be discussed in our  subsequent work.

\section*{Acknowledgments}
	 Authors thank the  Inter-University Centre for Astronomy and Astrophysics (IUCAA),  Pune for the facilities where this work was partially done under the visiting associateship programme.

\end{document}